\documentclass{article}
\usepackage[latin1]{inputenc}

\usepackage[usenames,dvipsnames,svgnames,table]{xcolor}
\usepackage{listings} 
\usepackage{url}
\usepackage[colorlinks=true,citecolor=Green]{hyperref}
\usepackage{epsfig}
\usepackage{fixltx2e}
\usepackage{balance}
\usepackage{colortbl}
\usepackage{multirow}
\usepackage{arydshln}
\usepackage{authblk}

\usepackage[margin=1in]{geometry}

\usepackage{marvosym}
\newcommand{\envelope}{(\raisebox{-.5pt}{\scalebox{1.45}{\Letter}}\kern-1.7pt)}

\lstdefinelanguage[ARM]{Assembler}%
  {morekeywords=[1]{.text,.globl,.align},%
  morekeywords=[2]{add, adr, sub, addc, adcs, adds, b, stmdb, %
    sub.W, beq,bge,blt,bne,%
    umull, umlal, cmp, ldr, mov, mul, pop, push, mrs, msr, %
    subs,vadd,vld1,vldm,vmov,vmul,%
    vpadd,vpop,vpush, str, b, bl,%
    itttt, ittee, itt, itte, it???, it??, itee, ittt, addne, subeq, moveq, subeq, movne, eorne, eor, rrx, 
   },%
   morekeywords=[3]{f32,s32,i32},
   morekeywords=[4]{r1,r2,r3,r4,r5,r6,r7,r8,r9,r10,r11,r12, lr, sp, apsr, rx, ry, rt, rhi, rlo, rz, rn, rm},
   keywordsprefix=.,%
   sensitive=false,%
   morecomment=[l]{;},
   moredelim=*[directive]\#,%
   moredirectives={define,elif,else,endif,error,if,ifdef,ifndef,line,%
      include,pragma,undef,warning}%
  }[keywords,comments,directives]

\definecolor{listing_background}{RGB}{250,250,250}
\definecolor{registers}{rgb}{0,0.4,0}
\definecolor{comments}{rgb}{0.4,0.4,0.4}

\title{Formal verification of a software countermeasure against instruction skip attacks}

\author[1,2]{Nicolas Moro}
\author[1]{Karine Heydemann}
\author[1]{Emmanuelle Encrenaz}
\author[2]{Bruno Robisson}

\affil[1]{Sorbonne Universités, UPMC Univ Paris 06, UMR 7606, LIP6, 75005 Paris, France \texttt{firstname.lastname@lip6.fr}}
\affil[2]{CEA, CEA-Tech PACA, LSAS, 13541 Gardanne, France ~~~~~~ \texttt{firstname.lastname@cea.fr}}

\hypersetup{
    pdftitle = {Formal verification of a software countermeasure against instruction skip attacks},
    pdfauthor = {Nicolas Moro, Karine Heydemann, Emmanuelle Encrenaz, Bruno Robisson},
    pdfsubject = {PROOFS 2013},
    pdfkeywords = {microcontroller, fault attack, instruction skip, countermeasure, formal verification}
}

\begin{document}\sloppy

\maketitle

\begin{abstract}
Fault attacks against embedded circuits enabled to define many new attack paths against secure circuits. Every attack path relies on a specific fault model which defines the type of faults that the attacker can perform. On embedded processors, a fault model consisting in an assembly instruction skip can be very useful for an attacker and has been obtained by using several fault injection means. To avoid this threat, some countermeasure schemes which rely on temporal redundancy have been proposed. Nevertheless, double fault injection in a long enough time interval is practical and can bypass those countermeasure schemes. Some fine-grained countermeasure schemes have also been proposed for specific instructions. However, to the best of our knowledge, no approach that enables to secure a generic assembly program in order to make it fault-tolerant to instruction skip attacks has been formally proven yet. In this paper, we provide a fault-tolerant replacement sequence for almost all the instructions of the Thumb-2 instruction set and provide a formal verification for this fault tolerance. This simple transformation enables to add a reasonably good security level to an embedded program and makes practical fault injection attacks much harder to achieve.
\end{abstract}

\makeatletter
\def\blfootnote{\xdef\@thefnmark{}\@footnotetext}
\makeatother

\blfootnote{The final publication is available at Springer via http://dx.doi.org/10.1007/s13389-014-0077-7}


\section{Introduction}
Physical attacks were introduced in the late 1990s as a new way to break cryptosystems. Unlike classical cryptanalysis, they use some weaknesses in the cryptosystems' implementations as a way to break them. Among them, faults attacks were introduced in 1997 by Boneh \textit{et al.} \cite{Boneh1997}. In this class of attacks, attackers try to modify a circuit's environment in order to change its behaviour or induce faults into its computations \cite{BarEl2006,Barenghi2012Book}. This attack principle was first introduced against cryptographic circuits but can be used against a larger set of embedded circuits. Many physical means can be used to induce such faults: laser shots \cite{Trichina2010}, clock glitches \cite{Balasch2011}, chip underpowering \cite{Zussa2012}, temperature increase \cite{Skorobogatov2009} or electromagnetic glitches \cite{Dehbaoui2012}.

Among fault attacks, three subclasses can be distinguished: differential fault analysis, safe error and algorithm modifications. Differential fault analysis (DFA) aims at retrieving some ciphering keys by comparing correct ciphertexts with ciphertexts obtained from a faulted encryption \cite{Boneh1997}. Safe-error attacks are based on the fact that a fault injection may have or not have an impact on the output \cite{Barenghi2012}. Finally, algorithm modifications target an embedded processor and aim at injecting faults into an embedded program's control flow \cite{Schmidt2008,Balasch2011}.

Those attack schemes rely on an attacker's fault model which defines the set of faults an attacker can perform \cite{Barenghi2012}. As a consequence, countermeasure schemes must take this fault model aspect into account. On microcontrollers and embedded processors, the fault model in which an attacker can skip an assembly instruction has been observed on different architectures \cite{Schmidt2008,Barenghi2012} and for different fault injection means \cite{Balasch2011,Dehbaoui2012,Trichina2010}. As a consequence, this fault model is a realistic threat for an embedded program.

In this paper, we consider this instruction skip fault model and propose a countermeasure scheme that enables to secure any assembly code against instruction skip faults. Some countermeasures based on multiple executions of a function have already been proposed and can theoretically handle this issue \cite{BarEl2006}. However, this kind of high granularity temporal redundancy is vulnerable to multiple fault attacks. Even with commonly-used low-cost fault injection means, a high temporal accuracy can be obtained by an attacker, and performing the same fault injection on several executions of an algorithm is practical \cite{Trichina2010}. On the contrary, performing faults on two instructions separated by a few clock cycles is significantly harder \cite{Barenghi2010} while still possible. Indeed, it requires a much more costly fault injection equipment and very high synchronization capabilities. It is then not yet considered as a realistic threat. 

The securing approach proposed in this paper uses an instruction-scale temporal redundancy to ensure a fault-tolerant execution of an embedded program. It is based on the statement that performing two faults on two instructions separated by few clock cycles is hardly feasible. 
 A fault-tolerant replacement sequence for most of the instructions of the whole Thumb-2 instruction set has been designed. We also show how to formally prove the fault tolerance of replacement sequences by using a model-checking tool.

By using such a fine-grained redundancy scheme, it is then possible to strengthen most assembly programs against fault attacks without any specific knowledge about the program itself. In the experimental results, we evaluate the overhead induced by fault tolerance and show that it can be reduced by only applying this countermeasure scheme to the sensitive parts of an implementation.

The rest of this paper is organized as follows. Section \ref{Section:RelatedWorks} introduces our fault model and  gives details about some related research papers. Section \ref{Section:ContreMesure} introduces our countermeasure scheme and details our replacement sequences. Section \ref{Section:PreuveFormelle} explains the approach we use for the formal verification. Finally, section \ref{Section:Implementations} evaluates the efficiency of our countermeasure scheme on several implementations. 


%

\section{Related works and fault model}
\label{Section:RelatedWorks}

This section is dedicated to related works. First, fault models are discussed in section~\ref{Paragraphe:InstructionSkips}. Countermeasure schemes that have previously been proposed are addressed in section~\ref{Paragraphe:Countermeasures}. Section \ref{Paragraphe:FormalVerification} presents some related research papers on formal verification. 

\subsection{Fault model}
\label{Paragraphe:InstructionSkips}

On embedded processors, a fault model in which an attacker can skip an assembly instruction or equivalently replace it by a \texttt{nop} has been observed on several architectures and for several fault injection means \cite{Karaklajic2009}. On a 8-bit AVR microcontroller, Schmidt \textit{et al.} \cite{Schmidt2008} and Balasch \textit{et al.} \cite{Balasch2011} obtained instruction skip effects by using clock glitches. Dehbaoui \textit{et al.} obtained the same kind of effects on another 8-bit AVR microcontroller by using electromagnetic glitches \cite{Dehbaoui2012}. On a 32-bit ARM9 processor, Barenghi \textit{et al.} obtained some instruction skip effects by using voltage glitches. On a more recent 32-bit ARM Cortex-M3 processor, Trichina \textit{et al.} were able to perform instruction skips by using laser shots \cite{Trichina2010}. Moreover, this fault model has also been used as a basis for several cryptanalytic attacks \cite{Barenghi2012}. As a consequence, it is considered as a common fault model an attacker may be able to perform \cite{Karaklajic2009}. 

A more generic fault model is the instruction replacement model, in which \texttt{nop} replacements correspond to one possible case. In some previous experiments on an ARM Cortex-M3 processor by using electromagnetic glitches, we have observed a corruption of the instructions binary encodings during the bus transfers \cite{Moro2013} leading to such instruction replacements. Actually, instruction skips correspond to specific cases of instruction replacements: replacing an instruction by another one that does not affect any useful register has the same effect as a \texttt{nop} replacement and so is equivalent to an instruction skip. Many injection means enable to perform instruction replacement attacks \cite{Moro2013,Balasch2011,Barenghi2012Book}. Nevertheless, even with very accurate fault injection means, being able to precisely control an instruction replacement is a very tough task and, to the best of our knowledge, no practical attack based on such a fault model has been published yet. 


As a conclusion, we consider in this paper the potentially harmful fault model in which an attacker is able to skip a single instruction.

\subsection{Countermeasure schemes}
\label{Paragraphe:Countermeasures}


Several countermeasures schemes have been defined to protect embedded processor architectures against specific fault models. At hardware level, many countermeasures have been proposed. As an example, Nguyen \textit{et al.} \cite{Nguyen2011} propose to use integrity checks to ensure that no instruction replacement took place. 

Software-only countermeasure schemes, which aim at protecting the assembly code, are more flexible and avoid any modification of the hardware. Against fault attacks, the most common software fault detection approach relies on function-level temporal redundancy \cite{BarEl2006}. For example, this principle applied to a cryptographic implementation can be achieved by calling twice the same encryption algorithm on the same input and then comparing the outputs. For encryption algorithms, an alternative way is to call the deciphering algorithm on the output of an encryption and to compare its output with the initial input. These approaches enable fault detection and involves doubling the execution time of the algorithm. Triplication approaches with voting enabling fault tolerance at the price of tripling the execution time of the whole algorithm have also been proposed~\cite{BarEl2006}. 

At algorithm level, in \cite{Medwed2008}, Medwed \textit{et al.} propose a generic approach based on the use of specific algebraic structures named \textit{AN+B} codes. Their approach enables to protect both the control and data flow. 

At assembly level, in \cite{Barenghi2010}, Barenghi \textit{et al.} propose three countermeasure schemes based on instruction duplication, instruction triplication and parity checking. Their approach ensures a fault detection for a small number of instructions against instruction skip or transient data corruption fault models. Our scheme enables a fault tolerance only against the instruction skip fault model but for almost all the instructions of the considered instruction set. Moreover, our countermeasure scheme has been formally proven fault tolerant. 

\subsection{Formal verification of software countermeasures}
\label{Paragraphe:FormalVerification}

Formal methods and formal verification tools have been used for cryptographic protocols' verification of to check that an implementation could meet the Common Criteria security specifications \cite{Chetali2008}. However, to the best of our knowledge, very few formal verification approaches to check the correctness of software countermeasure schemes against fault attacks have been proposed yet. One of the most significant contributions has been proposed by Christofi \textit{et al.} \cite{Christofi2013}. Their approach aims at performing a source code level verification of the effectiveness of a countermeasure scheme on a CRT-RSA implementation by using the Frama-C program analyzer. In this paper, we formally prove all our proposed countermeasures against an instruction skip fault model at assembly level. Another more recent contribution of a formal methodology at algorithm level has been proposed by Rauzy \textit{et al.} \cite{Rauzy2013}. In their scheme, an attacker can induce faults in the data flow of a target implementation described in a high-level language. This scheme enables them to detect unnecessary countermeasures or possible flaws on several CRT-RSA implementations.

\section{Countermeasure scheme}
\label{Section:ContreMesure}

The proposed countermeasure scheme aims at ensuring a fault-tolerant execution of an assembly code against instruction skip faults. The approach we propose relies on providing a formally proven fault-tolerant replacement sequence for almost all the assembly instructions of a whole instruction set. We chose the ARM Thumb-2 instruction set \cite{Thumb2} since ARM is a widely used target architecture for embedded processors. In this section, we give some details about the considered instruction set and present some of the replacement sequences we have defined for each instruction. This fine-grained redundancy scheme enables to strengthen most assembly codes against fault attacks without any specific knowledge about them.



\begin{table*}
\centering
\small
\caption{Instruction classes in the Thumb-2 instruction set}
\begin{tabular}{|l|l|l|} 
   \hline
   \rowcolor[gray]{0.92}\textbf{Instruction class} & \textbf{Examples} & \textbf{Replacement scheme} \\
   \hline
   Idempotent instructions & \texttt{mov r1,r8} & Instruction duplication \\
    & \texttt{add r3,r1,r2} & \\
   \hline
   Separable instructions & \texttt{add r1,r1,\#{}1} & Use of extra registers and \\
    & \texttt{push \{r4,r5,r6\}} & decomposition into an idempotent instruction sequence \\
   \hline
   Specific instructions & \texttt{bl $<$function$>$} & Replacement sequence specific to each instruction \\
    & \texttt{it} blocks & \\
   \hline     
\end{tabular}
\label{Tableau:ClassesInstructions}
\end{table*}

\subsection{The Thumb-2 instruction set}
\label{Paragraphe:Thumb2}
Thumb-2 is actually the successor to both ARM and Thumb instruction sets. Thumb-2 is a variable-length instruction set since it extends the 16-bit Thumb instruction set with some 32-bit instructions. Thus, 16 and 32-bit instructions can be mixed in a single program to combine both code density and performance. However, unlike the ARM instruction set, most 32-bit Thumb-2 instruction do not support direct conditional execution. To achieve such a conditional execution, a new \textit{If-Then} (\texttt{it}) instruction has been introduced. Moreover, some constant shifts can be applied to one operand of some instructions. Those shifts are : \texttt{lsl} (logical shift left), \texttt{lsr} (logical shift right), \texttt{asr} (arithmetic shift right), \texttt{ror} (rotate right) and \texttt{rrx} (rotate right one bit with extend). The Thumb-2 instruction set was first introduced with the ARMv6-T2 architecture and is now the standard instruction set for ARMv7 architectures. 

\subsection{Instruction classes}

We have defined a fault-tolerant replacement sequence for each instruction and each encoding of the Thumb-2 instruction set. This instruction set contains 151 instructions, and each instruction has up to four different encodings. For many instructions, the replacement sequence is very simple. However, this sequence can become much more complex for some specific instructions. According to the replacement sequences found, the instructions in the Thumb-2 instruction set can be divided into three classes. Every class is associated to one kind of replacement sequence. These three classes are summarized in Table \ref{Tableau:ClassesInstructions}.



The first class is composed of idempotent instructions which only need to be duplicated to provide fault tolerance. The second class gathers the instructions that are not idempotent but can be replaced by an equivalent sequence of idempotent instructions. The third class gathers some specific instructions that cannot easily be replaced by a list of idempotent instructions but for which a specific replacement sequence is possible. This last class also contains the instructions for which no replacement sequence that ensures fault tolerance and correct execution in any case can be provided. The solution for these instructions is either to avoid the compiler to use them or to use a fault detection approach. The following section gives more details about those classes. Moreover, it provides some examples of replacement sequences for every class.

%
%

\begin{table}[!h]
\small
\centering
\caption{Replacement sequences for some idempotent instructions}
\begin{tabular}{|l|l|l|} 
    \hline
    \rowcolor[gray]{0.92}\textbf{Instruction} & \textbf{Description} & \textbf{Replacement} \\
   	\hline
    \texttt{mov r1,r8} & Copies \texttt{r8} into \texttt{r1} & \texttt{mov r1,r8} \\
    &  & \texttt{mov r1,r8} \\
    \hline
    \texttt{ldr r1,[r8,r2]} & Loads the value & \texttt{ldr r1,[r8,r2]} \\
    & at the address & \texttt{ldr r1,[r8,r2]} \\
    & \texttt{r8+r2} into \texttt{r1} & \\
    \hline
    \texttt{str r3,[r2,\#10]} & Stores \texttt{r3} at & \texttt{str r3,[r2,\#10]} \\
    & the address \texttt{r2+10} & \texttt{str r3,[r2,\#10]} \\
    \hline
    \texttt{add r3,r1,r2} & Puts \texttt{r1+r2} into \texttt{r3} & \texttt{add r3,r1,r2} \\
    & & \texttt{add r3,r1,r2} \\
    \hline
\end{tabular}
\label{Tableau:InstructionsIdempotentes}
\end{table}

\subsection{Individual instruction replacement sequences}

\subsubsection{Idempotent instructions}
Idempotent instructions are the instructions that have the same effect when executed once or several times. If all the source operands are different from the destination operands, and if the value written into the destination operands does not depend on the instruction's location in the code, then the instruction is said to be idempotent. For such instructions, the countermeasure consists in a simple instruction duplication. The overhead for such a duplication is twofold: an overhead which equals the instruction size in terms of code size and a performance overhead that is equal to the execution time of the instruction. Table \ref{Tableau:InstructionsIdempotentes} gives some examples of idempotent instructions and their associated replacement sequence.

\subsubsection{Separable instructions}

\begin{lstlisting}[language={[ARM]Assembler},caption={Replacement sequence for the non idempotent \texttt{add r1, r1, r3} instruction\label{listing:separable_add}} ]
; we assume rx is an available register
mov	rx, r1
mov	rx, r1
add r1, rx, r3
add r1, rx, r3
\end{lstlisting}

In the considered instruction set, some instructions are not idempotent but can be rewritten by a sequence of idempotent instructions whose execution gives the same result. Once this rewritting is performed, each idempotent instruction of the replacement sequence can then be duplicated. This class gathers the instructions whose destination register is also a source register. To replace these instructions by a sequence of idempotent instructions, some extra registers have to be used. These registers have to be available at this location in the code: any dead register can be used\footnote{It turns out that, in the ARM calling conventions, the \texttt{r12} register can be used to hold intermediate values and does not need to be saved on the stack. Thus, this register can be used, if available, as a temporary register for such replacement scenarios.}. If no dead register is available, the stack can be used to temporarily store the value in a register.

\paragraph{Simple separable instructions}
Listing \ref{listing:separable_add} shows the replacement sequence for an \texttt{add r1, r1, r3} instruction. For this class of instructions, the overhead cost brought by our countermeasure scheme depends on the instruction to replace. There is an overhead cost in code size, performance and register pressure (since the replacement sequence needs some extra registers). For the \texttt{add r1, r1, r3} instruction example, one extra register is needed. Moreover, 4 instructions are required instead of 1 and the overhead cost in terms of code size is between 6 and 10 bytes (depending on the encoding used for the initial and the replacement instructions).

\paragraph{Stack manipulation instructions}
Some memory access instructions can update the address register before or after (\texttt{stmdb}\footnote{\texttt{stmdb} stores multiple registers into the memory and decrements the address before each access}/\texttt{ldmia}\footnote{\texttt{ldmia} loads a memory segment into multiple registers and increments the address after each access}) a memory access. As a consequence, this address register is both a source and a destination register for such an instruction. This is notably the case of the stack manipulation instructions (\texttt{push} and \texttt{pop}). These instructions respectively write or read on the stack and decrement or increment the stack pointer. Such instructions can be separated into a sequence of instructions that only perform one operation at a time, either a memory access or an address register update. The \texttt{push} instruction can be decomposed into instructions that first write the register to save on the stack and then decrement the stack pointer. As decrementing the stack pointer implies reading and writing the same register, this operation is decomposed into two steps in order to get a sequence of idempotent instructions. Such a replacement sequence for the \texttt{push} instruction is detailed on Listing \ref{listing:push}. This replacement requires 1 extra register and has a code size and performance overhead of 5 instructions.

\begin{lstlisting}[language={[ARM]Assembler},caption={Replacement sequence for the \texttt{push \{r1, r2, r3, lr\}} instruction\label{listing:push}}]
; the push{} instruction is equivalent
; to the stmdb sp!,{} instruction
stmdb	sp, {r1, r2, r3, lr}
stmdb	sp, {r1, r2, r3, lr}
sub	  rx, sp, #16
sub   rx, sp, #16
mov	  sp, rx
mov	  sp, rx
\end{lstlisting}

\paragraph{\texttt{umlal} instruction}
The \texttt{umlal} instruction multiplies two source registers and then adds the content of the concatenation of the two 32-bit destination registers. The final result is written into two 32-bit destination registers. As a consequence, this instruction has registers that are both source and destination. However, it can be decomposed. First, a multiply instruction whose result is a 64-bit value can be performed. Then the 64-bit addition has to be decomposed into several instructions. This requires to propagate the carry set by adding the 32 least significant bits (by using an \texttt{adds} instruction) to the addition of the 32 most significant bits by using an \texttt{adc} instruction. However the \texttt{adds} instruction sets the flags whereas the \texttt{umlal} does not: this sequence of instructions is not strictly equivalent to the \texttt{umlal} instruction and may be wrong if the flags are used after the \texttt{umlal} instruction without being set. As a consequence, it is necessary to save the flags before the sequence and restore them afterwards. Performing such a saving requires 4 extra instructions. The corresponding replacement sequence for this instruction is given in Listing \ref{listing:full-umlal}. This countermeasure requires 4 extra registers and replaces the initial instruction by 14 instructions. This replacement sequence is actually the most costly one of the whole instruction set, both in term of extra registers and extra instructions.

\begin{lstlisting}[language={[ARM]Assembler},caption={Replacement sequence for \texttt{umlal rlo, rhi, rn, rm } instruction that performs \texttt{rhi:rlo = rn*rm + rhi:rlo} \label{listing:full-umlal}}]
mrs   rt, apsr   ; save flags
mrs   rt, apsr
umull rx, ry, rn, rm
umull rx, ry, rn, rm
adds  rz, rx, rlo
adds  rz, rx, rlo
addc  rx, ry, rhi
addc  rx, ry, rhi
mov   rlo, rz
mov   rlo, rz
mov   rhi, rx
mov   rhi, rx
msr   apsr, rt  ; restore flags
msr   apsr, rt
\end{lstlisting}

\paragraph{Instructions with a constant shift}
As mentioned in Sec.~\ref{Paragraphe:Thumb2}, several constant shifts can be applied to one source operand of some instructions. Among them, the \texttt{rrx} shift rotates all the bits of the shifted register to the right by 1 and uses the carry to set the most significant bit. The carry is read by a \texttt{rrx} operation. Thus, if the initial instruction also writes the flags, it has to be decomposed. An example of replacement sequence for a \texttt{subs r1,r2,r3,rrx} instruction is provided in Listing \ref{listing:subs_rrx}.

\begin{lstlisting}[language={[ARM]Assembler},caption={Replacement sequence for a \texttt{subs r1, r2, r3, rrx} instruction\label{listing:subs_rrx}}]
rrx  ry, r3
rrx	 ry, r3
subs r1, r2, ry
subs r1, r2, ry
\end{lstlisting}

\subsubsection{Specific instructions}
\label{section:specific}
Some instructions cannot easily be replaced by a list of idempotent instructions. These instructions can still be decomposed into an equivalent sequence of instructions that can be duplicated to enforce a robust execution. There are also some instructions for which no fault-tolerant countermeasure in any case can be found. Some of them can still be replaced by a fault-tolerant sequence under some constraints. In this section, we give details and provide some examples for both kinds of such specific instructions.


\paragraph{\texttt{bl} subroutine call instruction}
The subroutine call instruction (\texttt{bl}) performs a jump and writes the return address into the link register (\texttt{r14}). Duplicating a \texttt{bl} instruction would induce two subroutine calls if no attack is performed. A solution is to explicitly put the return address into the link register and then perform an unconditional jump. As the Thumb execution mode requires the last bit of an instruction address to be set, this bit must be set before the unconditional jump to the subroutine code, as shown on Listing \ref{listing:bl}.

\begin{lstlisting}[language={[ARM]Assembler},caption={Replacement sequence for a \texttt{bl} instruction\label{listing:bl}}]
adr	ry,<return_label>
adr	ry,<return_label>
add lr, ry, #1 ; Thumb mode requires the
add lr, ry, #1 ; last bit of lr to be set
b	<function> 
b	<function>
return_label

\end{lstlisting}

\paragraph{Instructions that both read and write the flags}
Instructions that read and write the flags cannot easily be replaced by a fault-tolerant sequence of instructions. For example, the \texttt{adsc} instruction performs an addition between two source operands (two registers or one register and an immediate value) and the carry flag. The result is written into a destination register and the flags (carry, negative, overflow and zero) are updated. Duplicating such an instruction is not correct since the second \texttt{adcs} would use the carry set by the first \texttt{adcs} instruction instead of the initial carry value. If the flags are alive\footnote{A register is alive at a given point in an instructions sequence if there is a path to the end in which it is read before being written} after the \texttt{adcs} instruction, then no simple replacement sequence seems possible, the code has to be modified. Otherwise, if the flags are not alive after the \texttt{adcs} instruction, a replacement sequence exists. Such a sequence consists in saving the flags values before the first \texttt{adcs} instruction and restoring these values before the second \texttt{adcs} instruction. This replacement sequence is illustrated in Listing~\ref{listing:adcs}. 

\begin{lstlisting}[language={[ARM]Assembler},caption={Replacement sequence for a \texttt{adcs r1, r2, r3} instruction\label{listing:adcs}}]
mrs   rx, apsr  ; save flags
mrs   rx, apsr
adcs  r1, r2, r3
msr   apsr, rx  ; restore flags
msr   apsr, rx
adcs  r1, r2, r3
\end{lstlisting}

\paragraph{\texttt{it} blocks}
Thumb-2 provides conditional execution of instructions through \texttt{it} blocks. An \texttt{it} instruction specifies a condition and up to the 4 following instructions can be conditionally executed according to this condition or its inverse. \texttt{it} blocks correspond to if-then or if-then-else higher-level constructions and are useful when the branches of a conditional statement are composed of a limited number of instructions. Listing~\ref{listing:it_block} gives an example of such an \texttt{it} block. The simplest solution for such blocks is to first transform the \texttt{it} block into an equivalent classical if-then-else structure such as the one presented on Listing~\ref{listing:it_eq_cm} and then apply the countermeasure scheme to each instruction, as illustrated on Listing~\ref{listing:it_block2}.

\begin{lstlisting}[language={[ARM]Assembler},caption={Example of \texttt{it} block\label{listing:it_block}}]
  itte NE 
  addne r1, r2, #10
  eorne r3, r5, r1  
  moveq r3, #10 
\end{lstlisting}
\begin{lstlisting}[language={[ARM]Assembler},caption={Code equivalent to the \texttt{it} block of Listing~\ref{listing:it_block}\label{listing:it_eq_cm}} ]
  b.eq else
  add r1, r2, #10
  eor r3, r5, r1
  b continuation
else
  mov r3, #10
continuation
\end{lstlisting}%

 \begin{lstlisting}[language={[ARM]Assembler},caption={Code of Listing~\ref{listing:it_eq_cm} strengthened with individual instruction countermeasure scheme\label{listing:it_block2}} ]
  b.eq else
  b.eq else
  add r1, r2, #10
  add r1, r2, #10
  eor r3, r5, r1
  eor r3, r5, r1
  b continuation
  b continuation
else
  mov r3, #10
  mov r3, #10
continuation
\end{lstlisting}

\begin{table*}
\small
\centering
\caption{Summary of the defined instruction classes}
\begin{tabular}{|l|l|l|} 
   \hline
   \rowcolor[gray]{0.92}\textbf{Class} & \textbf{Type of instructions} & \textbf{Example} \\
   \hline
   Idempotent 
    & ALU operations & \texttt{add r1,r2,r3} - \texttt{subs r1,r2,\#{}8} \\
   \cdashline{2-3}[1pt/1pt]
   instructions
		
	& Load instructions & \texttt{ldrh r1,[r2,r3]} - \texttt{ldrb r1,[r2,\#{}8]} \\
   \cdashline{2-3}[1pt/1pt]
	& Store instructions & \texttt{strh r1,[r2,r3]} - \texttt{strb r1,[r2,\#{}8]} \\
   \cdashline{2-3}[1pt/1pt]
	& Branch instructions & \texttt{b <label>} - \texttt{bx lr} \\
   \cdashline{2-3}[1pt/1pt]	
	& Comparison instructions & \texttt{cmp r1,\#{}9} - \texttt{cne r3,r4} \\
   \hline
   Separable
    & Instructions with a register both source and destination & \texttt{add r1,r1,\#{}1} - \texttt{str r0,[r0,\#{}0]} \\
   \cdashline{2-3}[1pt/1pt]	
	instructions & Instructions with pre-indexed or post-indexed addressing & \texttt{str r1,[r2,\#{}8]!} - \texttt{str r1,[r2],\#{}8} \\ 
	\cdashline{2-3}[1pt/1pt]	
    & Stack manipulation instructions & \texttt{push \{r1,r2\}} - \texttt{pop \{r1,r2\}}  \\
    \cdashline{2-3}[1pt/1pt]	  
    & Instructions that use a \texttt{rrx} shift and write the flags & \texttt{adds r1,r2,r3,rrx} \\     
   \hline
   Specific & Subroutine call instruction &\texttt{bl $<$function$>$}  \\
   \cdashline{2-3}[1pt/1pt]	
   instructions & If-Then (\texttt{it}) blocks & \texttt{itte NE}  \\
   \cdashline{2-3}[1pt/1pt]	
   & Instructions for synchronization with external systems & \texttt{sev} - \texttt{yield} - \texttt{svc}\\
   \cdashline{2-3}[1pt/1pt]	
   & Instructions for coprocessors & \texttt{mcr p0,\#{}0,r8,r2,r3} \\
   \cdashline{2-3}[1pt/1pt]	
   & Instructions that both read and write the flags & \texttt{adcs r1,r2,r3} - \texttt{rrxs r1,r2} \\ 
   \hline
\end{tabular}
\label{Tableau:BilanClassesInstructions}
\end{table*}

\paragraph{Other replacement sequence for \texttt{it} blocks}
We have also designed a specific replacement sequence for \texttt{it} blocks but this replacement has some limitations and can quickly become more costly than its equivalent form with an if-then-else structure. Listing~\ref{listing:it_block} gives an example of such an \texttt{it} block. If the condition NE holds, (\textit{i.e. } if the Z flag is set), then the two following instructions (\texttt{addne} and \texttt{eorne}) are executed. Otherwise, the last two instructions (\texttt{subeq} and \texttt{moveq}) are executed. The whole \texttt{it} block needs to be considered in order to secure it. The solution we propose is to first apply our countermeasure scheme to every instruction of the \texttt{it} block. Every instruction of a replacement sequence first keeps the same condition as the initial instruction. The first \texttt{it} instruction is then duplicated. The second \texttt{it} instructions specifies one instruction less than the first one. Moreover, both \texttt{it} instructions are to be updated depending on the instructions that result of the replacement sequence of the instuctions of the initial \texttt{it} block. This step is presented in Listing~\ref{listing:cm1}. The second step consists in adding some \texttt{it} instructions, since \texttt{it} blocks cannot contain more than 4 instructions, as illustrated in Listing~\ref{listing:cm2}. Finally, the conditions set in the \texttt{it} instructions need to be updated to match with the instructions of the \texttt{it} block they define. Listing~\ref{listing:cm3} shows the secure code corresponding to the \texttt{it} block code example given in Listing~\ref{listing:it_block}.

\begin{lstlisting}[language={[ARM]Assembler},caption={First step for replacement of the \texttt{it} block given in Listing~\ref{listing:it_block}\label{listing:cm1}}]
it??? NE
it??  NE
addne r1, r2, #10
addne r1, r2, #10
eorne r3, r5, r1  
eorne r3, r5, r1  
moveq r3, #10 
moveq r3, #10 
\end{lstlisting}%

\begin{lstlisting}[language={[ARM]Assembler},caption={Second step for replacement of the \texttt{it} block given in Listing~\ref{listing:it_block}\label{listing:cm2}} ]
it??? NE
it??  NE
addne r1, r2, #10
addne r1, r2, #10
eorne r3, r5, r1  
it??? NE  
it??  NE
eorne r3, r5, r1  
moveq r3, #10 
moveq r3, #10 
\end{lstlisting}

 \begin{lstlisting}[language={[ARM]Assembler},caption={Final replacement sequence of the \texttt{it} block given in Listing~\ref{listing:it_block} \label{listing:cm3}} ]
itttt NE
ittt  NE
addne r1, r2, #10
addne r1, r2, #10
eorne r3, r5, r1  
ittee NE  
itee  NE
eorne r3, r5, r1  
moveq r3, #10 
moveq r3, #10 
\end{lstlisting}

Note that an \texttt{it} instruction should not appear in an \texttt{it} block. Thus, in case of a fault targeting one of the duplicated \texttt{it} instructions, the code behaves as if there was only one \texttt{it} instruction. Otherwise, the second \texttt{it} instruction is executed in the \texttt{it} block defined by the first \texttt{it} instruction. The second \texttt{it} instruction has actually no effect and is considered as a \texttt{nop}. However, some compilers may not accept such a construction. In this case, we have to use traditional conditional sequences for if-then or if-then-else constructions and apply our countermeasure scheme to each individual instruction of the resulting code as presented in Section \ref{section:specific}. Moreover, transforming first the \texttt{it} block into a classical if-then-else structure and then applying the countermeasure scheme may induce a smaller overhead cost. Such a construction has been previously presented in Listings \ref{listing:it_eq_cm} and \ref{listing:it_block2}.

\paragraph{Other specific instructions}
For some very specific instructions, defining a replacement sequence cannot really be done. Those specific instructions include the group of instructions for coprocessors (\texttt{mcr}, \texttt{lcr}, ...) or the instructions for synchronization with external systems (\texttt{sev}, \texttt{yield}, ...). 

\subsection{Summary of the defined instruction classes}
To sum up, an overview of the defined instruction classes is shown on Table \ref{Tableau:BilanClassesInstructions}. This table shows the different types of instructions that are included in every class and provides a few examples for each type of instructions.

\section{Formal verification of correctness and fault tolerance}
\label{Section:PreuveFormelle}

In this section, we present how we formally prove the fault tolerance specification for the countermeasure replacement sequences presented in Section~\ref{Section:ContreMesure}. Details about the models used for the verification approach are presented in Section~\ref{Paragraphe:Modelisation} and verification examples for some replacement sequences are presented in Section~\ref{Paragraphe:ExemplePreuves}.

\subsection{State machine model and specification to prove}
\label{Paragraphe:Modelisation}
A program acts as the application of transformations of the values stored in the set of registers or in memory. Each instruction of the program acts like a function whose input is a configuration of registers and memory and produces a new configuration. The program can then be represented as a transition system whose states are configurations of registers and memory, and any transition mimics the state transformation induced by an individual instruction execution.

\subsubsection{Individual instruction model}
Instead of proving the fault tolerance for a complete program, our model checking approach consists in proving the fault tolerance for each replacement sequence proposed in our countermeasure scheme. Indeed, it is sufficient to certify that the output state (registers and memory configuration) after the replacement sequence execution (with or without a fault injection) is equivalent to the normal output state after the initial instruction execution. As this output state is also the input state for the following instruction, using such a verification approach certifies that the next instruction will start from the right configuration. Moreover, this enables to use model checking while avoiding state-explosion problem.

\subsubsection{State machine model}
As explained before, we can model the execution of a sequence of instructions by a transition system $TS$. We define this transition system as $TS=\{S, T, S_0, S_f, L\}$. $S$ is the set of states, $T$ the set of transitions $T: S\rightarrow S$, $S_0$ and $S_f$ are the subsets of $S$ which respectively gather the initial states and final states. The final states from $S_f$ are absorbing states. A state from $S$ is defined by the value of the different registers (from the set of registers $R$ which includes the program counter) and processor flags (from the set of flags $F$). Each transition from $T$ is defined by the effect of an instruction on the registers and processor flags. $L$ is a set of labels which correspond to the values the program counter can take. An example of such a transition system for the \texttt{add r1, r2, r3} instruction is shown in Fig. \ref{fig:systeme_transitions}. To prove that a countermeasure for an instruction $i$ is robust against a fault, we build two transition systems: one for the initial instruction $m(i)$ and another one for its strengthened replacement sequence $m_{cm}(i)$.

\begin{figure}
\begin{minipage}{.48\linewidth}
\begin{verbatim}
# input r2, r3, flags
# output r1, flags
pc_init  : add r1, r2, r3
pc_final : next_instruction
\end{verbatim}
\end{minipage} \hfill
 \begin{minipage}{.48\linewidth}
$pc\_init, pc\_final \in L $ \\
$(R,F)$ is the current state\\
$(R',F')$ is the next state\\
$t: (R,F)\rightarrow (R',F')$ with \\
$R.pc = pc\_init$ \\ 
$R'.r1 = R.r1 + R.r2$ \\
$R'.pc = pc\_final$ 
\end{minipage} 
\caption{Transition system for the \texttt{add r1, r2, r3} instruction}
\label{fig:systeme_transitions}
\end{figure}

\paragraph{Fault model}
In any transition system $m_{cm}(i)$, one instruction skip fault may occur. An instruction skip fault is modeled by a transition from a state to any following one. Such a faulty transition only modifies the program counter. We add to the whole transition system a skip instruction faulty transition between every pair of adjacent states. As we assume that only one skip instruction fault injection may occur, every fault transition is guarded with a boolean which identifies that a fault has already occurred.


\begin{figure}
\centering
\epsfig{figure=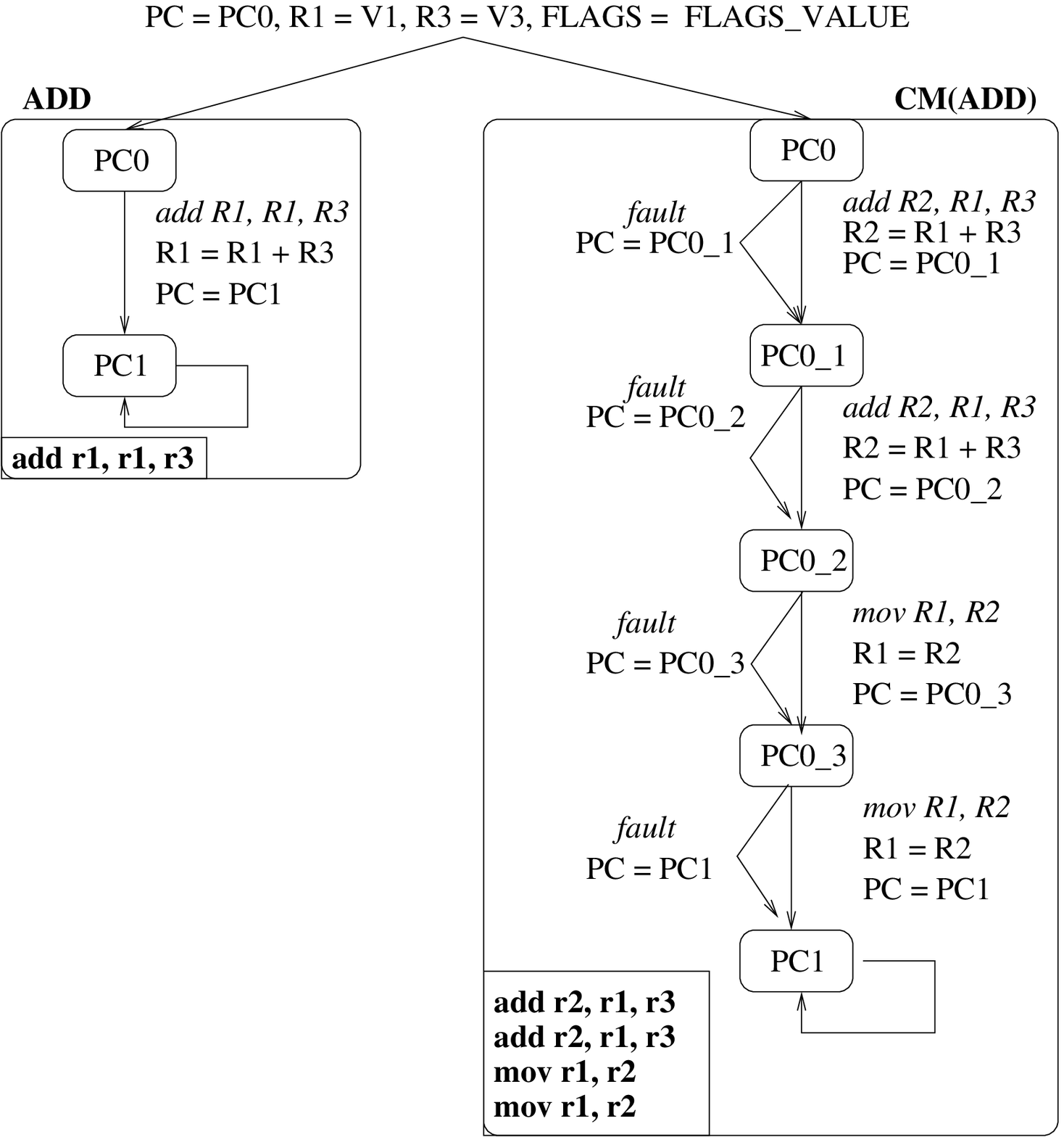,width=0.6\linewidth}
\begin{verbatim}
P1: AF(ADD.PC = PC1)
P2: AF(CM(ADD).PC = PC1)
P3: AG(((ADD.PC=PC1)*(CM(ADD).PC=PC1)) => 
     ADD.R1 = CM(ADD).R1 &
     ADD.FLAGS = CM(ADD).FLAGS)
\end{verbatim}
\normalsize
\caption{Model for a non-idempotent instance of the \texttt{add} instruction and its countermeasure}
\label{fig:model_add2}
\end{figure}

\paragraph{Flags and registers models}
The set of registers is composed of some general-purpose registers (\texttt{r0}-\texttt{r12}), a stack pointer (\texttt{r13}), a link register (\texttt{r14}) and a program counter (\texttt{r15}). The 5 processor flags are: C (carry), N (negative), Z (zero), V (overflow), Q (saturation). These flags can be set by some instructions and are used by several others. The conditional jumps are among the instructions that use those flags. Each flag is modeled as a 1-bit register. All the other registers are modeled as 4-bit registers. This width is sufficient to model the arithmetic and logic operations as well as the flags computations and enables to keep a reasonable complexity for the model checker. Moreover, modeling all the registers is not necessary since an instruction only reads a subset of the registers and writes on the destination registers. Besides, according to our fault model, the registers that are not modified by an instruction cannot be modified by a fault. Thus, for a given $m(i)$ or $m_{cm}(i)$, the set of registers $R$ is only composed of the subset of registers that are manipulated by $i$ or its replacement sequence $cm(i)$. Extra registers used in $cm(i)$ are supposed to be dead after the occurrence of the instruction $i$ in the initial program.

\paragraph{Memory model}
Since in our fault model we assume the memory cannot be corrupted, modeling the memory is not relevant. To ensure that a write to the memory took place, we only need to ensure that the corresponding instruction has been executed at least once. As explained later in this section, we add a counter variable to $m(i)$ and $m_{cm}(i)$ in order to achieve this. For the loads from the memory, we use symbolic values as the values cannot be corrupted and they also do not matter since the formal verification we use consists in checking the equivalence for any value. The important point is to give the same symbolic value to any loads at a given address for the transition system $m(i)$ (when $i$ is a \textit{load} instruction) and for $m_{cm}(i)$. This is achieved by adding an input variable for each memory address that is read by $i$ and $cm(i)$ to both transition systems. These variables contain the needed symbolic values.

\paragraph{Vis model checker}
We have chosen to use the Vis model checker\footnote{http://vlsi.colorado.edu/\~{}vis/} to prove the fault tolerance of our countermeasure scheme. This tool can take as input a transition system described with a subset of the Verilog hardware description language. Using Verilog is convenient to model transition systems which manipulate registers and bit vectors. The Vis model checker supports symbolic model checking techniques which enable to perform the proof in a symbolic way without having to enumerate each value for the registers. 
 In \cite{Fox2010}, Fox \textit{et al.} proposed a model of the ARMv7 architecture based on the HOL4 interactive theorem prover. However, this tool makes some proofs that are much more complex than the ones we need for the equivalence checking of finite models (the formal approach we propose to use in this paper) and the verification of rather simple properties on those models.


\begin{figure}
\centering
\includegraphics[width =0.6\linewidth]{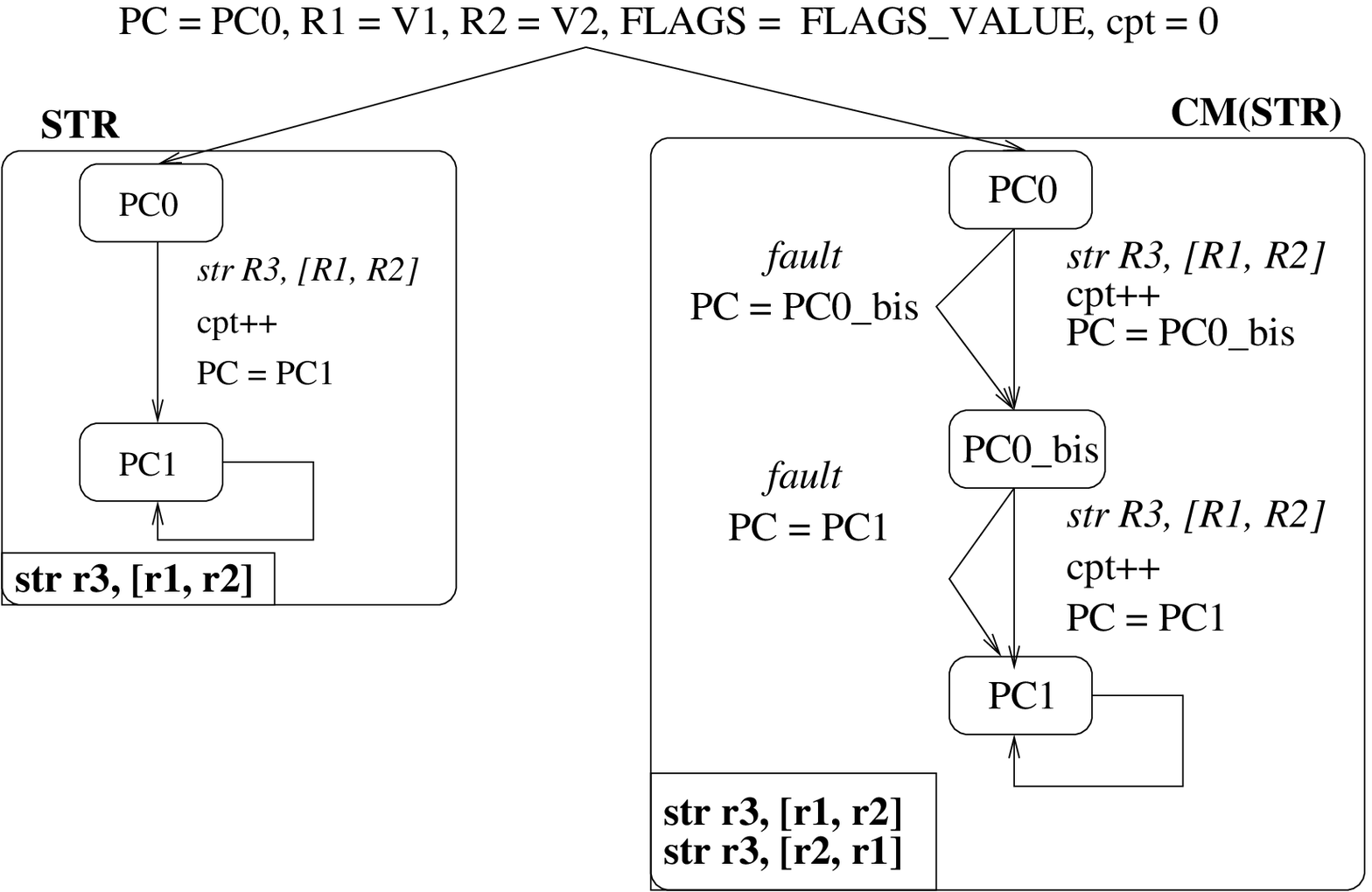}
\begin{verbatim}
P1 : AF(STR.PC=PC1)
P2 : AF(CM(STR).PC=PC1)
P3 : AG((STR.PC=PC1 * CM(STR).PC=PC1) =>
     (CM(STR).cpt = 2 + CM(STR).cpt = 1))
\end{verbatim}
\normalsize
\caption{Model for an idempotent instance of the \texttt{str} instruction and its countermeasure}
\label{fig:model_str}
\end{figure}

\subsubsection{Specification to prove}
\begin{figure*}
\begin{center}
\includegraphics[width =\linewidth]{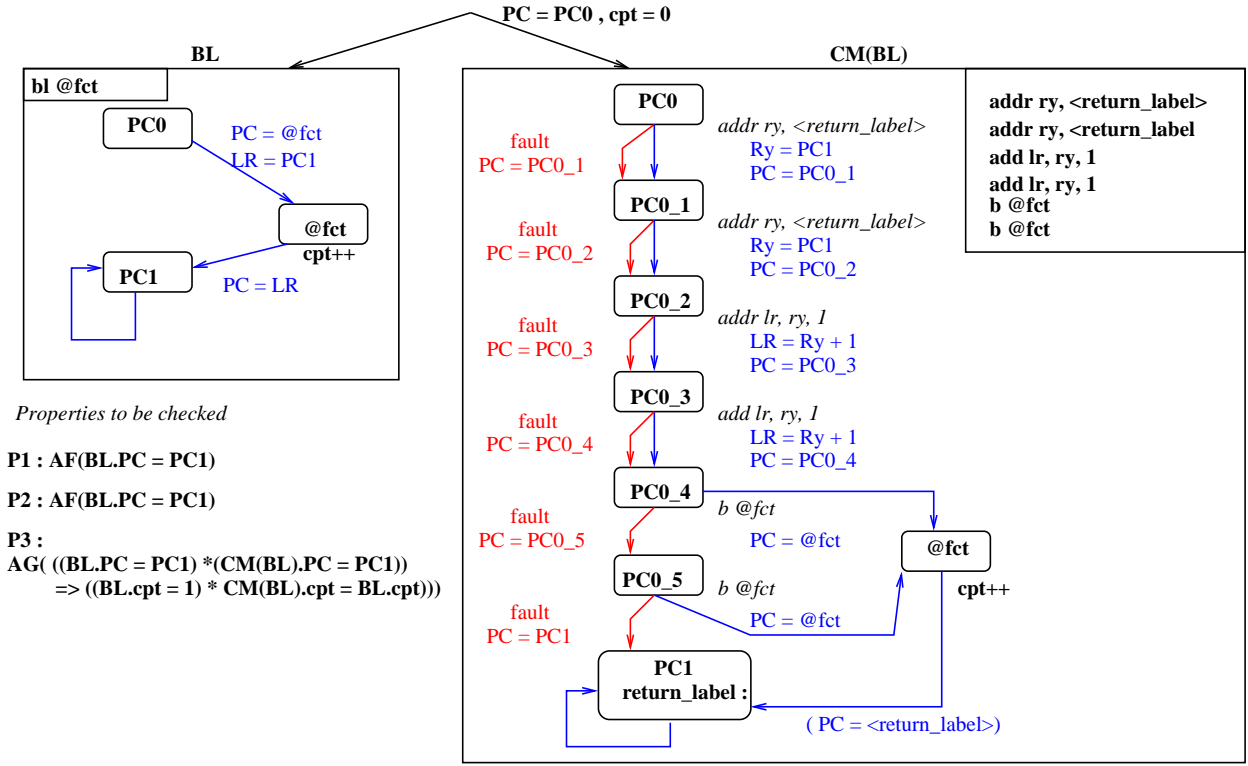}
\end{center}
\caption{Transition systems for the \texttt{bl} instruction and its replacement sequence}
\label{fig:model_bl}
\end{figure*}

To prove the equivalence of the output of an instruction and its replacement sequence, we prove the validity of logic formulas on the two models. To perform such a verification, we use a specific construction in which the two transition systems $m(i)$ and $m_{cm}(i)$ have the same values for the set of registers $R$ (except for the program counter), the set of flags $F$ and the symbolic values (for the memory loads) in their initial states. Such constructions are presented in Fig. \ref{fig:model_add2}, \ref{fig:model_str} and \ref{fig:model_bl}. We then need to prove that $m(i)$ and $m_{cm}(i)$ always reach a final absorbing state. Moreover, we also need to prove that, when $m(i)$ and $m_{cm}(i)$ reach a final state, the values for the set of alive registers $R'$ (except for the program counter) and flags $F'$ are similar. Such properties to check are expressed with the CTL temporal logic.

\begin{figure*}
\begin{verbatim}
MC: formula passed - AG(AF(adcs.pc=PC1)) 
MC: formula passed - AG(AF(cm(adcs).pc=PC1)) 
MC: formula passed - AG(((adcs.pc=PC1*cm(adcs).pc=PC1)->LIGHT_RESULT=1))
MC: formula failed - AG(((adcs.pc=PC1*cm(adcs).pc=PC1)->RESULT=1))
\end{verbatim}
\caption{VIS Model Checker output for the equivalence checking of the \texttt{adcs} instruction}
\label{fig:vis_output}
\end{figure*}

\subsection{Formal verification for some replacement sequences}
\label{Paragraphe:ExemplePreuves}

\subsubsection{Idempotent and separable instructions}
The left part of Fig.~\ref{fig:model_add2} shows the state machine corresponding to the transition system for a non-idempotent \texttt{add r1,r3} instruction. The program counter is updated and depending on the instruction, the registers or the flags may be updated too. The replacement sequence uses a dead register \texttt{r2} and two extra \texttt{mov} instructions to write the result to the destination register \texttt{r1}. Its transition system is modeled by the state machine on the right part of Fig.~\ref{fig:model_add2}. To prove that the replacement sequence is fault tolerant against a possible instruction skip, both state machines are fed with the same values for the source registers (\texttt{r1} and \texttt{r3}) and flags. Then, the validity of three CTL logic formulas has been checked with the Vis model checker. P1 and P2 express the fact that in both state machines any path from an initial state goes to a final state. P3 expresses the fact that in this final state, for all possible values in the source registers, the values in \texttt{r1} and the flags are identical in $m(i)$ and $m_{cm}(i)$. Fig.~\ref{fig:model_str} presents the transition systems for an idempotent memory write, namely an \texttt{str r3, [r1, r2]} instruction, and its replacement sequence. In this case, as the instruction writes the content of \texttt{r3} to the memory at the address \texttt{r1}+\texttt{r2}, and as we consider an instruction skip fault model, no proof is needed on the value hold by the registers. We only need to make sure that at least one \texttt{str} instruction has been executed. A counter variable is added to the definition of a state. This counter is set to 0 and is incremented by any transition which corresponds to a \texttt{str} instruction. P1 and P2 express the fact that any path goes to the last state. P3 expresses the fact that the number of writes made by the replacement sequence greater or equal to the number of writes made by the initial instruction (which is equal to 1).

\begin{table*}
\footnotesize
\centering
\caption{Countermeasures overhead for several implementations}

\begin{tabular}{|l|c|c|c|c|c|c|} 
   \hline
   \multirow{2}{*}{\textbf{Implementation}}  & \multicolumn{2}{c|}{\cellcolor[gray]{0.92}\textbf{Without countermeasure}} & \multicolumn{4}{c|}{\cellcolor[gray]{0.92}\textbf{With countermeasure}} \\
   \cline{2-7}
   & \textbf{Clock cycles} & \textbf{Code size} & \textbf{Clock cycles} & \textbf{Increase} & \textbf{Code size} & \textbf{Increase} \\
   \hline
   \textbf{AES}	& 9595 & 490 bytes & 20503 & 113.7 \% & 1480 bytes & 202 \% \\
   \hline
   \textbf{MiBench AES} & 9294 & 3372 bytes & 26618 & 186.4 \% & 9776 bytes & 189.9 \% \\
   \hline
   \textbf{MiBench SHA0} & 4738 & 746 bytes & 10558 & 122.8 \% & 2076 bytes & 178.2 \% \\
   \hline
   \textbf{AES with countermeasure}	& 9595 & 490 bytes & 11374 & 18.6 \% & 1874 bytes & 282.5 \% \\ 
   \textbf{on the last two rounds} & & & & & & \\
   \hline	   
\end{tabular}
\label{Tableau:Implementations-Overhead}
\end{table*}

\subsubsection{Specific instructions}
\paragraph{Subroutine call: the \texttt{bl} instruction}
Figure~\ref{fig:model_bl} shows the state machines for the \texttt{bl} instruction and its replacement sequence. In both corresponding transition systems, we have added a label $@fct$ to model the target of the subroutine call. Transitions from a state in which $PC=@fct$ assign the link register to the PC. Such a transition models the return of the function and also increments a counter. Then, properties P1 and P2 to be checked by the model checker express that any path from an initial state goes to a final state. Property P3 expresses the fact that in a final state the number of calls to the function (the counter values) are the same. Validity of property P3 ensures that the function has been executed only once while validity of P1 and P2 ensures that the control flow comes back to the calling function.

\paragraph{Instructions that read and write the flags}
For the \texttt{adcs} instruction and its replacement sequence, as presented in Listing~\ref{listing:adcs}, the CTL properties are the same as the ones that were used for the \texttt{add} instruction. However, the property that deals with the equality of the destination register and the flags is not valid if a fault targets the last \texttt{adcs} instruction. Relaxing the constraint on flags equality (expressed as \texttt{LIGHT\_{}RESULT}) makes this property valid as shown with the output of the Vis model checker in Fig.~\ref{fig:vis_output}. To sum up, this countermeasure can only be used if the flags are not used before being set again after the \texttt{adcs} instruction.

\subsubsection{Verification statistics}
The verification process is based on modeling the arithmetic and logic operation performed by instructions and modeling their effects on the flags. Thus, the minimal size required to model the registers is the minimal one that enables to precisely model the effects on the flags. As a consequence, modeling registers with a 4-bit size is relevant. However, the verification process leads to the same results (passed or failed) for registers modeled with a 4-bit length or for larger sizes. Moreover, the verification process duration and the size of the transition system are related to the size of the registers. The verification process required less than 1 second per instruction with a 4-bit register size for all the instructions, and less than 1 minute for almost all the instructions with a 16-bit register size. As an example, it needed 29 hours to complete for the most expensive \texttt{umlal} instruction with a 16-bit register size. Moreover, the model checker we used could not build the internal representation or carry out the verification for registers larger than 24 bits for all the instructions. 


\section{Application to several implementations}
\label{Section:Implementations}

In this section, we applied our countermeasure scheme to several codes. Two of them are implementations of the AES-128 symmetric encryption algorithm, and the last one is an implementation of the SHA-0 algorithm. We developed the first AES implementation, in which every round key is calculated before the associated \texttt{AddRoundKey} operation. The second AES implementation and the SHA-0 implementation come from the MiBench embedded benchmark suite~\cite{Guthaus2001}. We provide an estimation of the overhead cost brought by our countermeasure scheme for those three implementations and perform an exhaustive instruction skip simulation on an ARM Cortex-M3 microcontroller to confirm the effectiveness of our approach. The chosen target is an up-to-date 32-bit microcontroller based on the ARM Cortex-M3 processor \cite{DefinitiveGuideARMCortexM3}. This microcontroller uses an ARMv7-M Harvard architecture and runs the Thumb-2 instruction set \cite{Thumb2}. 

\subsubsection*{Estimation of the overhead cost}
The overhead cost in terms of clock cycles and code size for the three implementations that use our countermeasure scheme is shown on the first three lines of Table \ref{Tableau:Implementations-Overhead}. For those implementations, the whole code has been strengthened with our methodology and both overhead costs are high. 

Another approach consists in applying our countermeasure to a specific chosen part of the algorithm. As an example, in terms of cryptanalysis, fault injections are supposed to be harder to exploit if the fault does not target the last two rounds. Thus, as shown on the fourth line of Table \ref{Tableau:Implementations-Overhead}, it could be possible to reduce the overhead in terms of clock cycles by applying our countermeasure scheme to the last two rounds only. This last scenario is just a possible example of optimization and some cryptanalysis attacks may still exist. Nevertheless, it aims at showing that the overhead costs can be significantly reduced with a good knowledge about the algorithm to strengthen and about some possible vulnerabilities in its assembly code. It is important to mention that all the instructions from the tested codes could be replaced by using our countermeasure scheme. Moreover enough dead registers were always available for all the replacement sequences that required some extra registers.

To sum up, the overhead cost brought by our countermeasure scheme is high, but remains comparable to the one brought by classical algorithm triplication or other software approaches for fault tolerance. However, unlike such classical algorithm duplication or triplication approaches, our countermeasure scheme should be resistant to double fault attacks in a time interval longer than a few clock cycles.

\section{Conclusion}
In this paper, we have presented a countermeasure scheme that enables to strengthen an embedded program and make it tolerant to instruction skip faults. In our countermeasure scheme, we have built a fault-tolerant replacement sequence for almost all the instructions of the whole Thumb-2 instruction set. The instructions can be divided into three classes, which all have their dedicated replacement sequences. We have also provided a formal verification process in order to guarantee the correctness and the fault tolerance of our replacement sequences for each class of instructions.

Finally, we do not claim our scheme enables a full protection against fault attacks. Nevertheless, such an approach enables to add a reasonably good security level to an embedded program, without requiring any extra hardware countermeasure and any specific knowledge about the embedded program. The overhead cost brought by using such a countermeasure is comparable to the extra cost brought by using classical algorithm-level temporal redundancy approaches and can be reduced with a more accurate knowledge about the sensitive parts that should be protected. Moreover, using a very fine-grained redundancy at the instruction scale makes the multiple fault attacks less practical with a reasonable cost equipment.
	

\bibliographystyle{plain}
\bibliography{library}

\end{document}